%% file: rpis_report.tex
\documentclass[11pt,a4paper]{article}
\usepackage[utf8]{inputenc}
\usepackage[english]{babel}
\usepackage{amsmath}
\usepackage{amsfonts}
\usepackage{amssymb}
\usepackage{graphicx}
\usepackage{siunitx}
\usepackage{float}
\usepackage{xcolor}
\usepackage{listings}
\usepackage{breakcites}
\usepackage[backref=page]{hyperref}
\definecolor{ao}{rgb}{0.0, 0.5, 0.0}

\hypersetup{
    unicode=false,
    pdftoolbar=true,
    pdfmenubar=true,
    pdffitwindow=false,
    pdfnewwindow=true,
    colorlinks=true,
    linkcolor=blue,
    citecolor=blue,
    filecolor=magenta,
    urlcolor=ao
}

\usepackage[margin=2.5cm]{geometry}
\author{Daniel H. Biedermann$^{\mbox{\tiny{a}}}$\footnote{Corresponding author. Tel.: +49-89-289-25060; fax: +49-89-289-25051. \newline
\textit{E-mail address:} daniel.biedermann@tum.de}, Felix Dietrich$^{\mbox{\tiny{a,b}}}$, Oliver Handel$^{\mbox{\tiny{a}}}$, Peter M. Kielar$^{\mbox{\tiny{a}}}$, Michael Seitz$^{\mbox{\tiny{a,b}}}$\\\\
$^{\mbox{\tiny{a}}}$ Technische Universität München\\
$^{\mbox{\tiny{b}}}$ Hochschule München}

\date{}

\title{Using Raspberry Pi for scientific video observation of pedestrians during a music festival}
\begin{document}

\maketitle

\tableofcontents

\clearpage
\begin{abstract}
The document serves as a reference for researchers trying to capture a large portion of a mass event on video for several hours, while using a very limited budget.
\end{abstract}

\section{Introduction}

Pedestrians and crowds have become an increasingly researched subject. Many scientific disciplines are involved when studying people's behaviour, such as psychology, sociology, systems theory and various engineering disciplines. The principal practical motivation is the safety and comfort of pedestrians \cite{Handel2014}. However, other qualities, for example openness or profitability of events, must not be neglected in a holistic treatment of the matter.

Computer simulations can help to identify possible hazards and can therefore be used as a tool to support the planning of mass events. Pedestrian dynamics simulations can be grouped into different categories. Macroscopic models describe pedestrians not as individuals but as cumulated densities, and are often based on the theory of fluid dynamics (e.g. \cite{Lighthill1955, Richards1956, Hartmann2013}). They simulate fast, but with low spatial resolution \cite{Anh2012}. In contrast, microscopic models describe pedestrians as individual and discrete objects. Many different microscopic models have been proposed in literature to simulate pedestrians virtually (e.g. \cite{gipps-1985,helbing-1995,moussaid-2011, seitz-2012,dietrich-2014}, for a comprehensive review  see \cite{zheng-2009,duives-2013}). A third type of pedestrian simulation models are so called hybrid models, which combine different pedestrian dynamic models (e.g. \cite{Xiong2010, Anh2012, Biedermann2014}, for a comprehensive review see \cite{Ijaz2015}). Such multiscale approaches support a holistic view on public events \cite{Biedermann2014a}. For the simulations to produce credible outcomes, their models have to be calibrated and validated with empirical data. Data collection is challenging, because many factors that can influence the behaviour of a crowd have to be registered. Therefore, a key method is to video record the scene and analyzes the video footage later. This is a common approach to validate models  of pedestrian dynamics (e.g. \cite{Schadschneider2009,Davidich2013, Kielar2014576}).

This report describes the technical and practical application of Raspberry Pi computers \cite{upton-2014} in combination with a camera unit and dedicated Linux operating system for scientific video observation of an event. Different technological solutions for tracking pedestrians have been proposed \cite{lemercier-2011, dollar-2012, seer-2014}. Here we describe a low-budget approach using various Raspberry Pi units for video recording. The approach was used to observe a music festival in Bavaria, Germany, in 2014, with about 5000 visitors. The areas of interest for later analysis were preselected and the positions of the cameras were chosen to best cover them. The observation is part of the research project MultikOSi\footnote{Assistance systems for urban events -- multi criteria integration for openness and safety, \url{www.multikosi.de}}. 

In Section \ref{sec:case} we begin by describing the observed event as case study for the technical approach. In Section \ref{sec:material} we first give a detailed account of the materials and methods used in preparation of, during, and for processing the video footage after the event. In Section \ref{sec:safety}, we discuss data privacy and safety considerations, for both the visitors of the event and the scientific staff conducting this study. Finally, in Section \ref{sec:discussion}, we conclude, give an outlook of possible future work, and discuss some possible solutions to issues that occurred during the observation.

\section{Case study ``Back to the Woods’’}
\label{sec:case}
\input{Chapter_CaseStudyBTTW}

\section{Material and Methods}
\label{sec:material}

\subsection{Raspberry Pi based video recording system}
\label{sec:cammonitoring}
\input{./Chapter_CameraMonitoring.tex}

\subsection{Evaluation of the video results}
\input{./Chapter_MatlabTool.tex}

\section{Data privacy and safety considerations}
\label{sec:safety} 
\subsection{Data privacy considerations and procedures}
\input{./Chapter_DataPrivacy}

\subsection{Safety considerations for the research team}
\input{./Chapter_OnEventSafety}

\section{Discussion}
\label{sec:discussion}

In this Section we describe various issues that occurred during the study and suggest solutions. Afterwards, this leads to the conclusions of this paper and the assessment of the approach is presented. Finally, we give an outlook on future work and further use of the gathered data in the research project MultikOSi.

\subsection{Lessons learned}
The Raspberry Pi units can be complemented with an additional wireless LAN unit. This would allow better positioning of the hardware in preparation of the observation, because, without wireless LAN, the Raspberry Pi unit has to be connected to a laptop computer with a wire. Additionally, after setting up the hardware it can still be checked and controlled remotely, which is not possible if the unit is positioned at an inaccessible place, such as a tree.

The time should be noted when setting up the units at the beginning. This is necessary for aligning the video footage and data later.

The configuration of Raspberry Pi units and battery packs was successful. The batteries lasted the expected time and longer. Battery packs are highly advantageous compared to wired power supply. Wired powering brings various problems. First, positioning of the units is more difficult with a wire. Second, the power cord has to be placed and can be obstructive during the event. Third, a socket-outlet must be present or provided for every unit. Fourth, socket-outlets and wires represent a possible hazard to the visitors, event staff, and scientific staff, especially in wet weather conditions. Fifth, a power outage might stop video recording if wired powering is used.

Although the Raspberry Pi and camera units proved to be highly reliable, important areas that have to be observed should be covered with two units positioned at different places. Furthermore, cameras should be positioned at places that cannot be reached by visitors. In one case the alignment of a video recording changed after visitors climbed the stage where the units were placed. However, considering the rather low costs of the units, even the loss of a camera might be acceptable if the camera cannot be placed at a safer place and the recording is important.

The plastic boxes and the duct tape used to protect the units have proven to be efficient. No Raspberry Pi or battery pack was damaged during the observation, even though a strong thunder storm with heavy rain hit the festival. However, this will also depend on the carefulness the equipment is positioned and the boxes are sealed with.

Finally, the time necessary to set up the equipment was underestimated. Although five members of scientific staff worked on positioning and setting up the units, it took up to 45 minutes until the equipment was ready. This time may be reduced with experience, but generally should be planned with. Assigning specific roles to staff beforehand might reduce this problem.

\subsection{Conclusions}
Overall we deem the approach described in this report as very successful. The two main requirements were met: the equipment is cost effective and we obtained the desired outcome in form of video footage. Despite the adverse weather conditions no Raspberry Pi unit was damaged. The collected video footage is of sufficient quality for the analysis. The battery backs used for some cameras proved to be effective and showed several advantages over the wired powering. Wireless LAN units could furthermore facilitate the placement and control of the Raspberry Pi computers in the field.

However, some disadvantages can be identified. There are relatively high demands on technical expertise, such as at least a basic knowledge about the Linux operating system used. Furthermore, the preparation and placement is time consuming and has to be planned sufficiently ahead of the observation. Finally, the lightning conditions have to be good in order for the cameras to yield reasonably quality of video recordings. Unfortunately, the resolution of the video footage is not very high. However, the latter two points could be improved with different camera units for higher resolution.

\subsection{Outlook and future work}
The data collected from this observation is used for statistical analysis and model development. In a first step, positions of pedestrians were annotated and projected in two-dimensional world coordinates. This data yields anonymous but individual positions and trajectories of visitors. Based on such raw data further information about the crowd can be gained, such as the individual velocity or personal space in Voronoi diagrams \cite{steffen-2010,duives-2015}. Furthermore, the spatial distribution of visitors on the venue or the shape of queues in front of vendor stands and the entrance can be analysed over time.

Some research objectives using the data are: the service times in front of vendor stands; degree of capacity utilisation of toilets; safety considerations for visitors during the event; spatial distribution of visitors on the venue; shape of and individual behaviour of pedestrians in ; individual trajectories as input for an agent-based simulation model; calibration of microscopic pedestrian simulation models, such as the optimal steps model \cite{seitz-2012,seitz-2015}, the concurrent hierarchical finite state machine based pedestrian decision model \cite{Kielar2014576}  and the gradient navigation model \cite{dietrich-2014}.

A second observation of the same event is scheduled for the year 2015. For this observation, the research groups plan on implementing all lessons learned. Furthermore, the data collected from this second observation can then be used to validate the hypotheses and models developed based on the first observation.

Technical improvements are mainly the extension of the Raspberry Pi computers with wireless LAN controllers. Furthermore, the new version, Raspberry Pi 2, could be used. An alternative to the camera units could be infra-red cameras that would also have the advantages of anonymous recordings, although this could be less cost effective.

\section*{Acknowledgements}
We would like to thank Mr. Daake from the safety department, the commissioner of data protection Dr. Baumgartner, and the plant fire brigade of the Technical University Munich for their support. Furthermore, we would like to thank Mr. Bügler, whose large experience and expertise with the Raspberry Pi computer system was crucial and necessary for our successful observations.

This work was partially funded by the German Federal Ministry of Education and Research through the project MultikOSi on assistance systems for urban events -- multi criteria integration for openness and safety (Grant No. 13N12824). The authors gratefully acknowledge the support by the Faculty Graduate Center CeDoSIA of TUM Graduate School at Technische Universit\"{a}t M\"{u}nchen, Germany. Support from the TopMath Graduate Center of TUM Graduate School at Technische Universit\"{a}t M\"{u}nchen, Germany and, from the TopMath Program at the Elite Network of Bavaria is gratefully acknowledged

\bibliographystyle{plain}
\bibliography{rpis_report}

\end{document}

%% file: Chapter_CaseStudyBTTW.tex
In order to test the technical equipment in a real-world event context, in this Section we document a case study where the approach was used. The studied event took place at the campus Garching of the Technical University of Munich in 2014. The event is called ``Back to the Woods'' and had over 5000 visitors. In the following some basic information and benchmark data about the event is provided to the reader.

\subsection{Description of the event}
\begin{figure}
\centering
\includegraphics[width=0.8\textwidth]{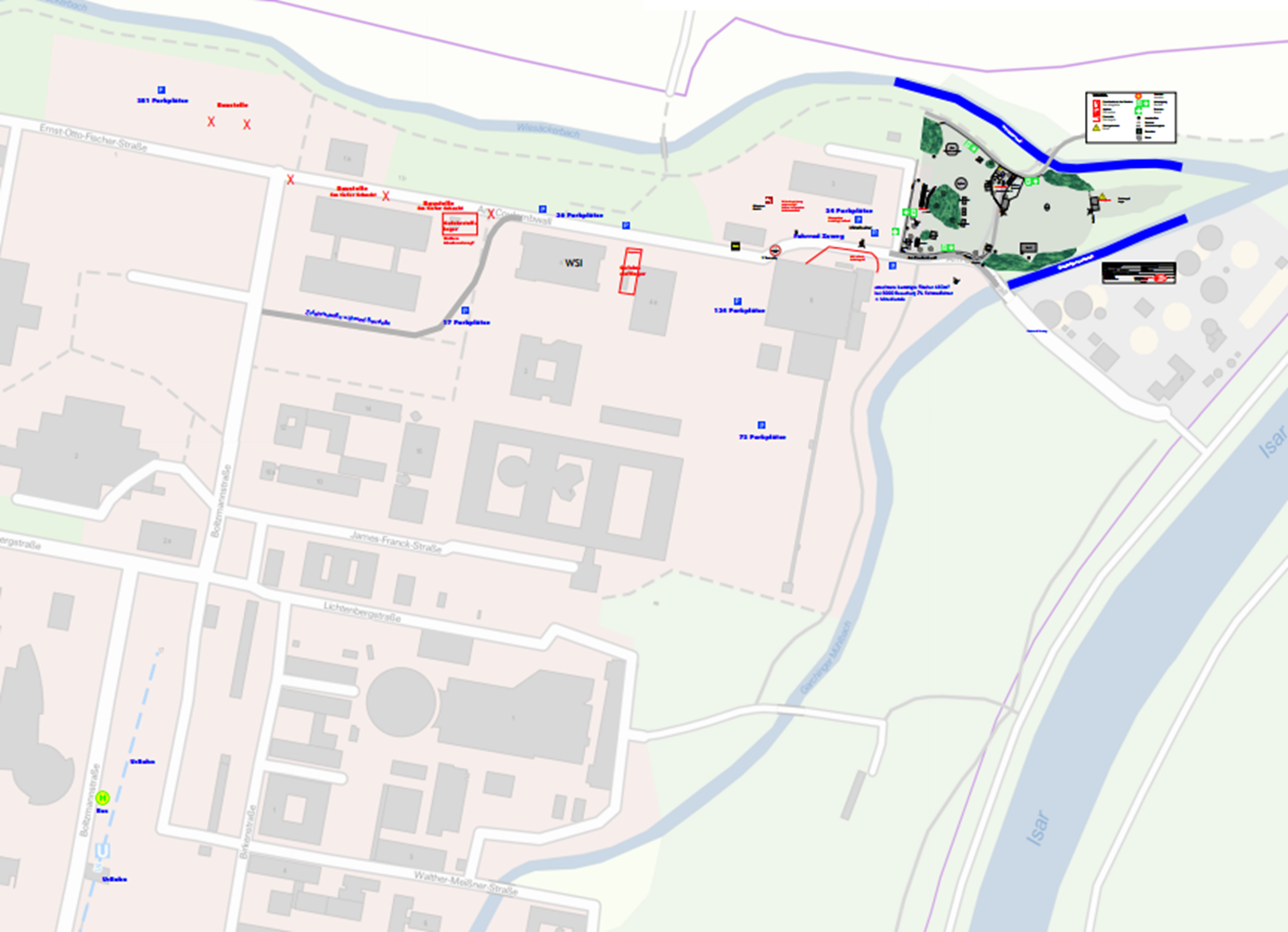}
\caption{\label{fig:3AreaMeso}Location of the event area at the campus Garching.}
\end{figure}

The event had the character of an open air festival and took place on July, the 27th in 2014. The festival area was situated at the edge of the university campus next to the Isar River on an approximately 9000 square meter large meadow (see Figure \ref{fig:3AreaMeso}). The area is surrounded in the north, east and south by two little creeks. From this perspective the area was at the end of a blind alley with a major entrance route coming from the west, which leads to the metro station Garching Forschungszentrum (placed at the bottom left in Figure \ref{fig:3AreaMeso}) from where the bulk of visitors arrived.

\begin{figure}
\centering
\includegraphics[width=0.8\textwidth]{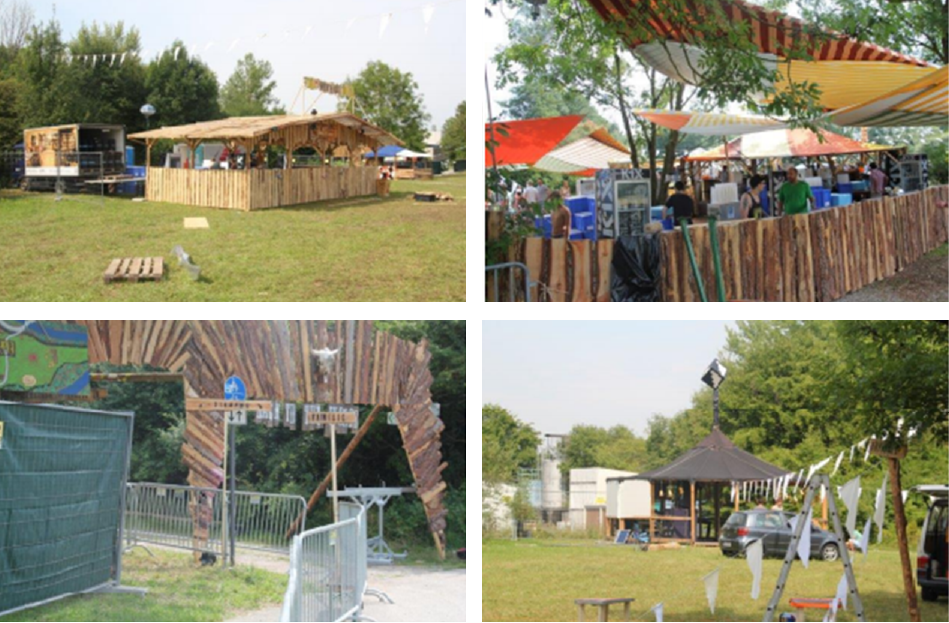}
\caption{\label{fig:EventImpressions}Impressions of the wooden facilities in the run-up of the event (taken by Michael \"{O}hlhorn, Vabeg® Eventsafety).}
\end{figure}

The festival consisted of two different public areas: one area with the main stage and two big bars and another area with a smaller DJ stage in a 60 square meter jute, food stands, toilettes, and a zone dedicated to relaxing. Another backstage area behind the main stage was used for equipment and artists. The infrastructure for facilities, electricity and music equipment was provided with temporary solutions (power generators, mobile toilettes, etc.). Apart from the toilettes, all facilities were made of wood and had a rustic-style in concert with the name of the event. The played music was electro, techno and rave. Figure \ref{fig:EventImpressions} shows some impressions of the facilities before the event had started.

\subsection{Camera positions}
\begin{figure}
\centering
\includegraphics[width=0.9\textwidth]{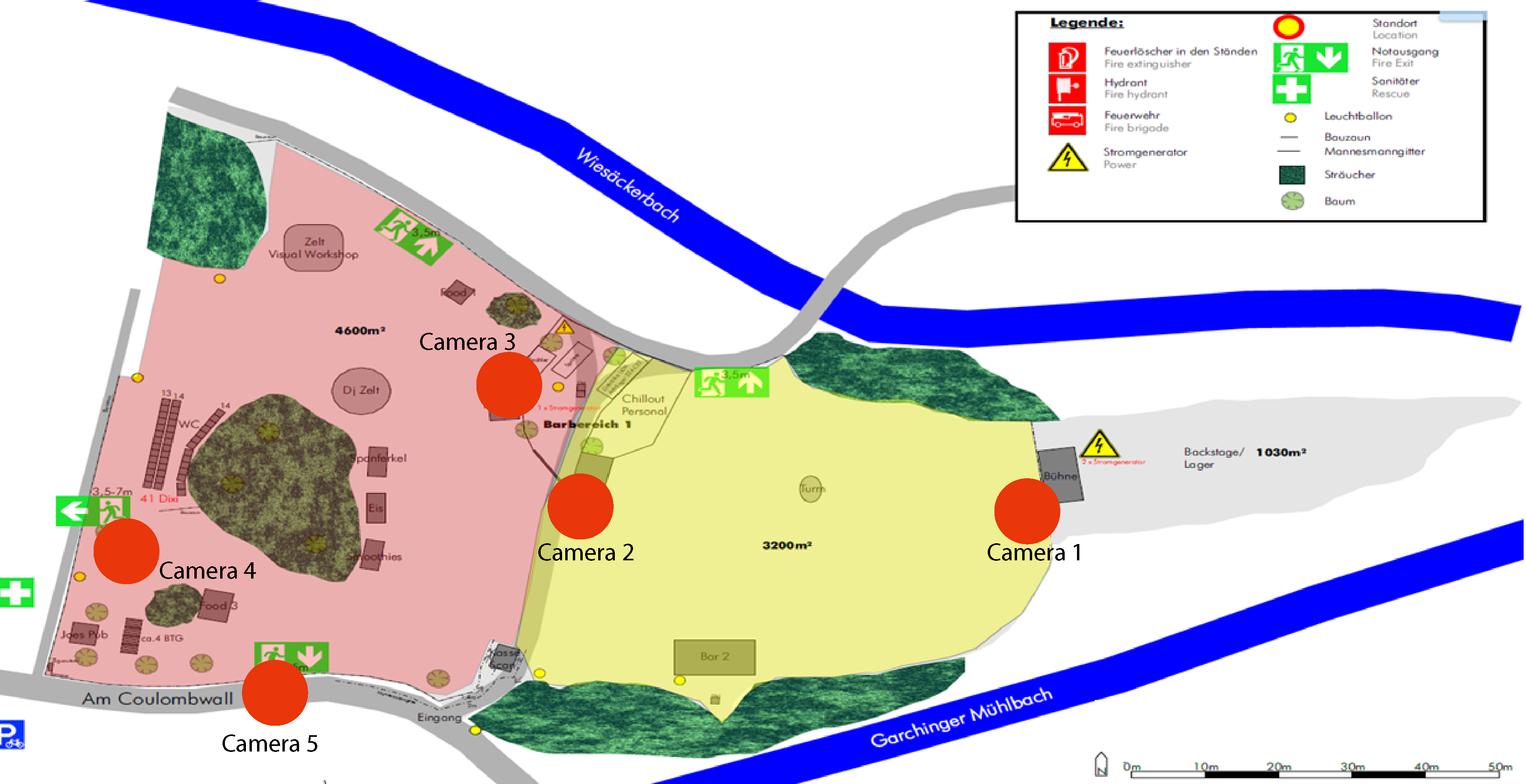}
\caption{\label{fig:CameraPositions}Layout plan of the festival with the positions of the camera recordings.}
\end{figure}

In total five different video recording units were used. In Figure \ref{fig:CameraPositions} the locations of the camera positions are shown.

\begin{figure}
\centering
\includegraphics[width=0.7\textwidth]{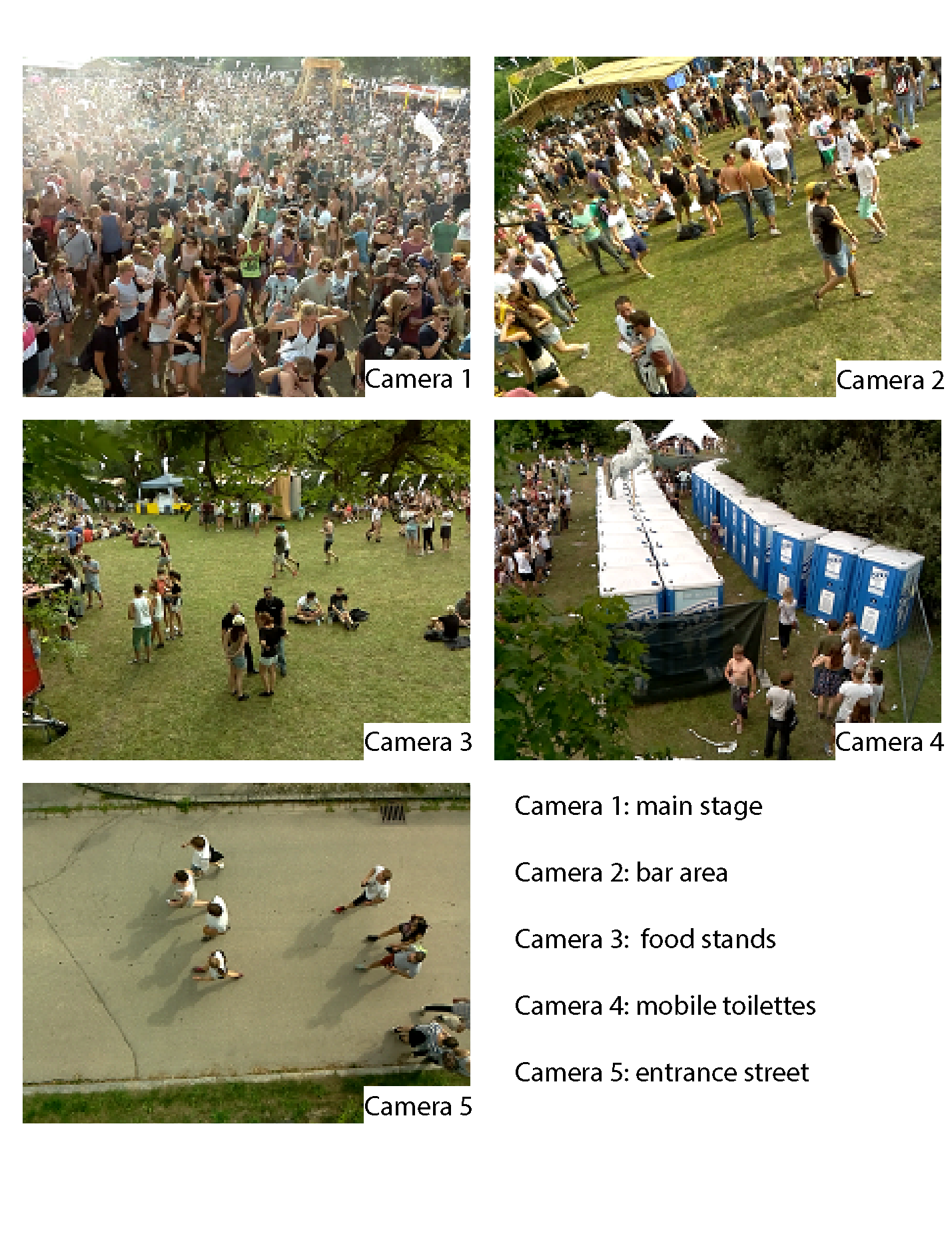}
\caption{\label{fig:3CameraScreenshots}Screenshots of the different camera recordings.}
\end{figure}

The first camera was placed on the main stage with a wide view over the crowd in front. The second camera was situated at the central bar with the view in the direction of the bar in the south. Camera number three was aligned to observe the three main food stands, to be able to evaluate their utilisation rate. The utilisation rate of the toilettes was observed by the fourth camera. The last camera filmed the profile of the major entrance street from a light post. In Figure \ref{fig:3CameraScreenshots}  screenshots of every stationed camera system are shown.
\subsection{Set up of the camera systems and issues to consider}
As suspension points for the camera systems, trees, event facilities and a light post were used. For hanging up the cameras a hydraulic lifting platform was used (see Figure \ref{fig:Safety2}). The set-up process of the Raspberry camera systems contained four different steps.
\begin{enumerate}
\item Start the video capturing mode with the help of a LAN cable and a laptop and record the starting time of the capturing.
\item Hang up the camera at the suspension point and mount the rain shelter.
\item Adjust the camera in order to capture the desired camera picture.
\item Arrange four traffic cones in a square at the ground of the camera picture to capture the alignment of the ground plane.
\end{enumerate}
\subsubsection{Issues}
The recording of all camera systems was successful. Nevertheless, three issues should be taken into consideration when working with the described Raspberry Pi camera system. First, as discussed in Section \ref{sec:cammonitoring}, the power for the camera systems was based on external Power Banks. Due to the long duration of the event, the power for the cameras did not last as long as  the festival. The second issue is related to the absence of a WLAN module at the Raspberry Pi computer. Without an external WLAN adapter, it is not possible to control the recording process without having a connected LAN cable. It seems to be helpful to add external USB WLAN controllers to be able to check the recording process from time to time. The third and last issue concerns the suspension points. It is cheap and easy to use existing suspension points such as trees, light posts and event facilities. In our context, one should prefer massive and difficult to access mounting options over those that are affected by wind or visitors climbing masts and thereby changing the view of the camera.
\subsubsection{Course of the event}
\begin{figure}
\centering
\includegraphics[width=0.7\textwidth]{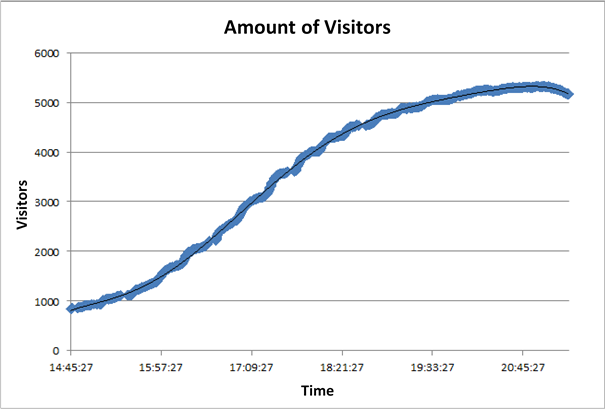}
\caption{\label{fig:VisitorsOverTime}Accumulated amount of people inside the festival area based on post evaluation of camera 5. The curve is S-shaped.}
\end{figure}

4500 tickets had been sold in advanced sales and could be printed by the visitors at home, or, alternatively, they could bring the ticket on their mobile phone to the event. Another approximately 500 tickets were sold at the event. In the entrance area a waiting line was formed out of crowd barriers. The entrance procedure included bag search, ticket control and putting on the entry wristband. The event started at 12 p.m. (noon), but approximately half of the visitors arrived between 4 at 6 p.m., the other half arrived before or after this time interval (see Figure \ref{fig:VisitorsOverTime}). In the Figure the peak number of visitors is about 5300, which exceeds the amount of 5000 available cards. This is can be explained by the amount of employees, visitors on the guest list and stowaways, which are not considered in this number.

\begin{figure}
\centering
\includegraphics[width=1\textwidth]{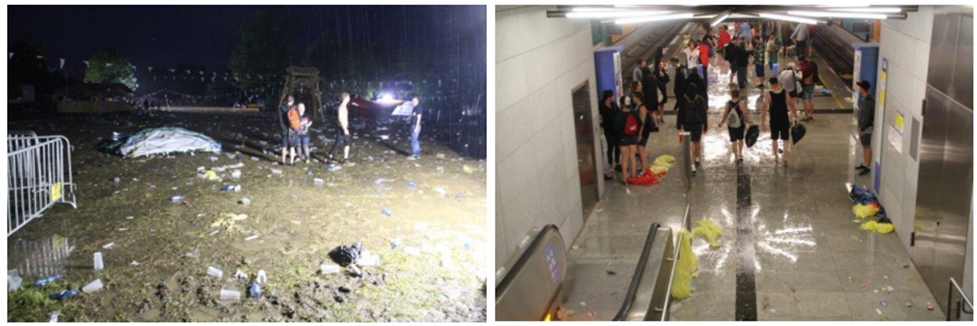}
\caption{\label{fig:StormImpressions}Left picture – Festival area after heavy rain falls. Right picture – Metro station Garching Forschungszentrum with leaving event visitors.}
\end{figure}

Besides long waiting times at the entrance (up to 45 minutes at the peak), at the toilettes (up to 10 minutes) and the bars plus a few ambulance operations due to high summer temperatures, no special occurrences happened until the evening. The planned end of the festival was scheduled for 2 a.m., but at 8 p.m. the weather service issued a storm warning.  Around 45 minutes later heavy rain and lightning strikes made it necessary to evacuate the festival area and to coordinate the pedestrian flows to the metro station. In Figure \ref{fig:StormImpressions}  two pictures are depicted from impressions during the storm. In the next Section, the evaluation of the camera observations are discussed. The focus is set to the run of the event and not to the evacuation procedure.

%% file: Chapter_CameraMonitoring.tex
\begin{figure}
\centering
\includegraphics[width=0.9\linewidth]{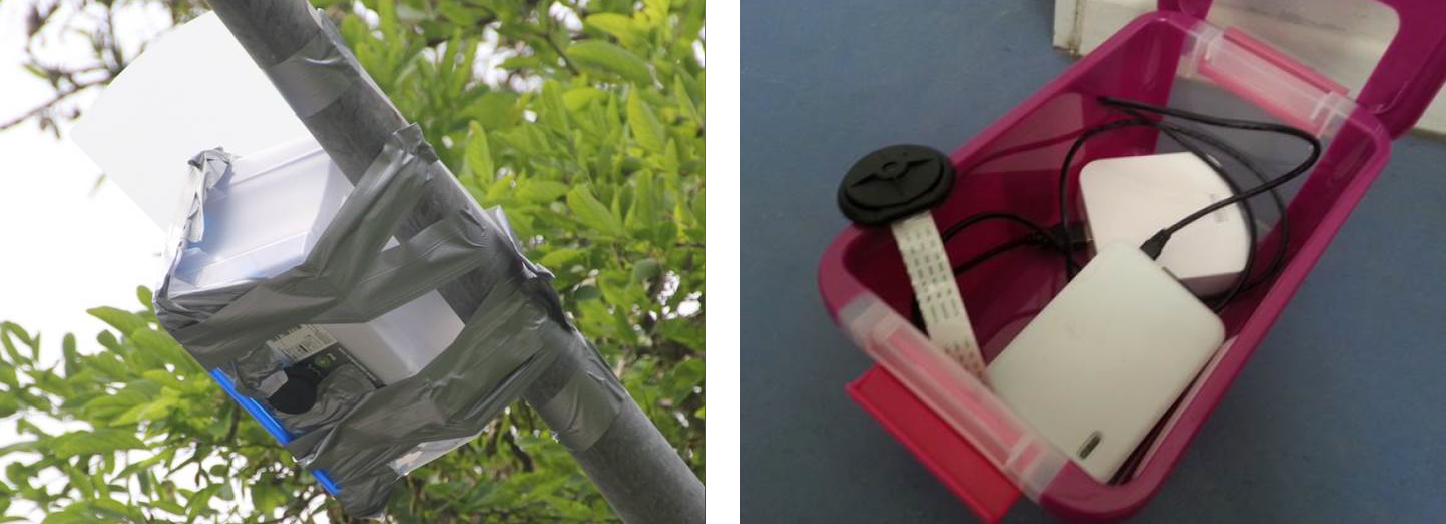}
\caption{The camera system and its singular components.}
\label{fig:cameras}
\end{figure}

A robust and automatic camera system was necessary to record the pedestrian flows during the observed music festival. The best solution to fulfil these restrictions was a self-compiled camera system. It is based on the microcomputer Raspberry Pi (RPi), which was extended by a camera module. A list of the used components can be found in table \ref{tab:RPIcomponents}. The camera system and the singular components are shown in Figure \ref{fig:cameras}.
 
Before the camera system can be used, an operating system has to be installed on the microcomputer. For video capturing, a stripped-down Linux distribution like Arch Linux is sufficient. Since the RPi does not have an internal hard drive, it has to be extended by an external one (e.g. a SD-card) with enough storage (at least 8 gigabyte) for the operating system. Another hard disk is mandatory to store the video footage.  An one gigabyte storage is sufficient for a recording about 8 minutes with 25 frames per second. Since most public events last no longer than 8 hours, a hard drive with 75 gigabyte of storage is adequate.

The RPi can be accessed with a standard SSH-Client. The access is possible by a local area network (LAN) with cable or wirelessly. In the context of public events, both methods have pros and cons. If the RPi is extended by a wireless adapter, the settings of the recording can be easily changed during the event. The problem here is that an open wireless connection can easily be accessed by unauthorised people. Therefore, mainly due to data protection, it is necessary to apply a secure network. This issue does not occur if a LAN cable is used. In this case, the camera systems can be placed on areas inaccessible for event visitors (e.g. high trees or buildings). This approach has two disadvantages. Firstly, the removal of the cable can influence the alignment of the camera module. And secondly, physical contact with the RPi is necessary. The camera system is normally placed on high and difficult to reach objects to record a large area of the event. The installation of the camera system normally takes place before the beginning of the observed event. But once it is started, it can be very challenging to reach the system due to the possibly large amount of visitors. So the change of camera settings during the event can be difficult to realize. In our case, we needed a large lifting platform to place the camera system on trees for the observed festival. This procedure would have been impossible if visitors had been near these areas. 

An  independent energy supply is needed to realize a self-sufficient camera system. We used Power Banks (battery chargers) as accumulators because of their USB-interface, which is ideal to connect the Power Bank with the used microcomputer. A battery with 10400\si{\milli\ampere\hour} will last about 8 hours, which should be sufficient for most public events. The benefits of a self-sufficient camera system could be proven on the studied event as a strong storm cut the energy supply of the festival. Without the batteries, no recordings could have been made after this incident.
 
The camera system was placed inside of a plastic case (circa 20\si{\cm}$\times$15\si{\cm}$\times$10\si{\cm}) to protect the electronic components. Possible negative environmental influences are rain or direct solar irradiation. The plastic box itself can be made waterproof by patching the case with duct tape. Concerns that a closed box could lead to an overheating of the RPi were unnecessary since the microcomputers worked on ambient temperatures over 30\si{\celsius} without any disruption. In spite of a strong storm during the  event, the recorded data and all cameras remained unscathed. This demonstrates that the RPi units can be effectively protected with simple plastic boxes during an open air festival.  
 
\begin{table}
\begin{center}
\begin{tabular}{|l|l|}
component & usage  \\ \hline
camera extension & video recording \\
Raspberry Pi motherboard& controlling of the camera module  \\ 
hard drives (USB / SD-card) & to save OS and recordings \\
battery & energy supply \\
box  & weather protection \\
\end{tabular} 
\end{center}
\caption{Singular components of the camera system}
\label{tab:RPIcomponents}
\end{table} 
The camera module can be controlled via the corresponding programme ``raspvid''. It is a command-line based application and can be easily run by bash-scripts. The Bash is a command processor, which is typically controlled by command lines or small scripts.  The following exemplary lines mount the hard drive, start the camera module, record a video with 25 frames per second for one minute with $1600\times 1200$ pixels and write them in the H.264 format to the mounted hard drive:
\begin{lstlisting}[language=bash]
#!/bin/bash
mount /dev/sda1 /mnt
screen -s video raspvid -fps 25 -t 60000 
       -w 1600 -h 1200 -o /mnt/Filename.h264
\end{lstlisting}
The command \textit{-fps} controls the number of frames, which get recorded per second. A larger frame rate increases the amount of memory necessary to film the complete event. Parameter \textit{-t} defines the duration of the recording in \si{\milli\second} and \textit{-w}/\textit{-h} describe the width/height of the recording in pixels. The parameter \textit{-o} defines the path and filename to the output file of the recording. A complete description of the ``raspvid'' can be found in the documentary \cite{DocumentaryRPi}. The current angle of vision of the camera can be tested in any browser by entering the local IP address of the RPi. This live preview has to be used with caution, because  the actual recording can be different if the recording's relation of width and height does not equal 4:3. The reason for this is that the preview uses the photography mode instead of the movie mode.

%% file: Chapter_MatlabTool.tex
Since most of the generated data from the event consists of video material, an important issue was the extraction of useful information out of the video data. It was accomplished by a tool written in MATLAB that enables the user to perform the following actions:
\begin{enumerate}
\item Load a specific video.
\item Jump to a given frame.
\item Mark any number of points of interest in the given frame.
\item Move the frame pointer back and forth 
\item Store the marked points of interest together with their respective time/frame index in a csv file.
\end{enumerate}
With this tool, it is possible to measure the density over time, or track individuals over some time interval. It has already been used successfully to evaluate a stepping experiment. Participants were filmed walking on a straight line. The video material captured was then used to compute the step lengths and speeds \cite{seitz-2014c}.

\begin{figure}
\centering
\includegraphics[width=0.8\textwidth]{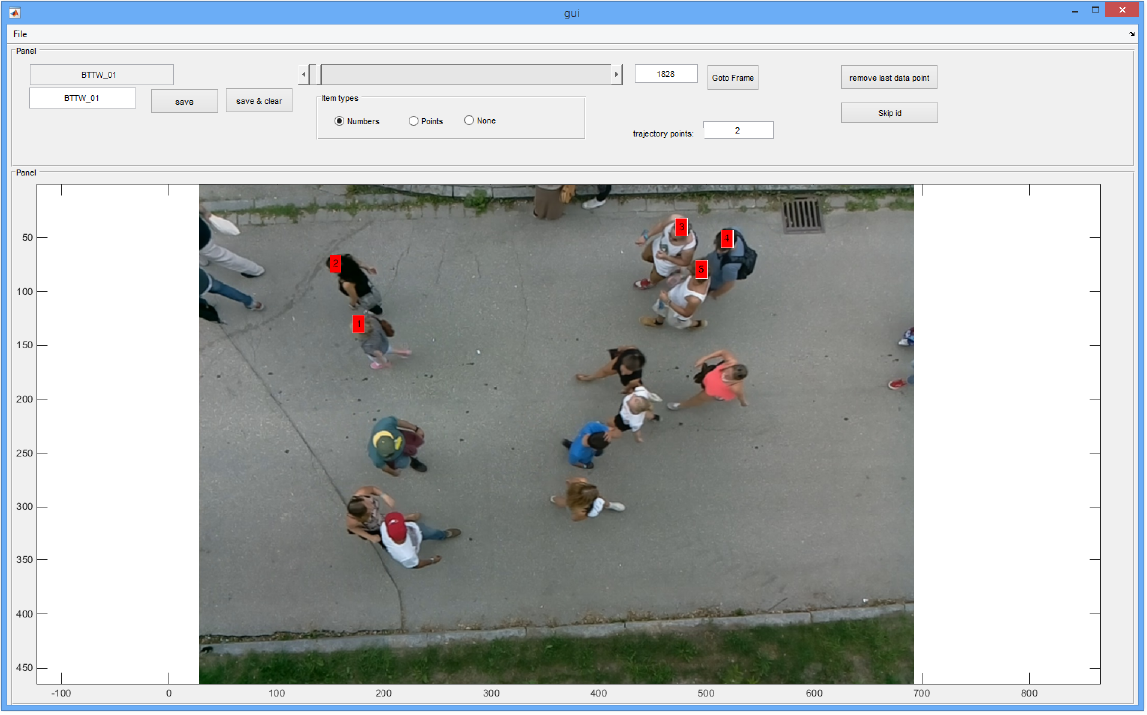}
\caption{\label{fig:MATLABGui}MATLAB tool used to evaluate the video material. This screenshot shows a frame from the camera on a lamp post facing downwards and people moving from left to right and from right to left. Five people in this frame are marked with red boxes; the tool automatically assigns numbers and stores the time index.}
\end{figure}

Figure \ref{fig:MATLABGui} shows a screenshot of the tool. The video captured by a camera on a lamp post facing downwards on the street is loaded. The frame shows 13 persons and some additional persons partly cut off. Five persons are already marked with the tool (red boxes); the numbering in the boxes is added automatically.

%% file: Chapter_DataPrivacy.tex
Camera observation is regulated by law in Germany. These laws regulate privacy and security issues of people who are or may be filmed by the cameras, and the data handling during and after surveillance. The research group had to comply with the Data Protection Acts in order to be legally allowed to use cameras for research. 

The main factor of uncertainty regarding compliance to these laws is that German Data Protection Act and the State Data Protection Acts differ in some cases regarding camera surveillance. The states' laws and the federal laws differ regarding content and interpretation. Even if the general legal basis is quite similar for all the different Data Protection Acts, we focused on the Bavarian Data Protection Act (Bayerisches Datenschutzgesetz, BayDSG) for handling the camera surveillance request for scientific research. A camera surveillance request can be granted for scientific research data acquisition at festivals like the ``Back To The Woods'', if a set of steps are carried out carefully.

 The key for understanding the steps is that for applying the BayDSG one cannot differentiate between a research group that wants to use camera surveillance for research purposes, and a company department that wants to use cameras for property safety and security issues. Therefore, the term \textit{company} is used to describe the organiser of the festival or the mainly responsible organisation of the festival, and the term \textit{department} is used for our research group.

 This Section summarises all mandatory steps for using camera surveillance in Bavaria (Germany) in order to respect the Data Protection Acts for research purposes. The summary can be applied to most research purposes and may be used as a starting point for other research teams interested in camera supported research.

 Five documents have to be prepared beforehand:
\begin{enumerate}
\item short review article
\item method description
\item description of the video recording devices
\item data handling description regarding video data processing, storage and deletion
\item staff council letter
\end{enumerate}
\subsubsection{Short review article}
The article 21a \cite{BayDSG} obliges that a short review article, comprised of details regarding legitimation of the video surveillance procedure for research purposes, is mandatory. The article helps the company to apply approval criteria for camera surveillance requests. The short review article is stored at the companies office and will be presented to the regulatory commissioner of data protection if demanded.
\subsubsection{Method description}
A company has to maintain a directory of all released camera surveillance methods within the companies jurisdiction regarding article 27 \cite{BayDSG}. Therefore, a method description, which will also be approved by the company, has to be prepared by the department. The method description is obliged by article 26 clause 3(1) \cite{BayDSG}. A method description starts with general information of the department. The first Section of the description comprises of the objective and the legal basis of the data acquisition. The next Section has to present a detailed description of the involving computational data processing, the data storage concept, the standard periods regarding data deletion, and the way the department will use the collected data. If the data is send to company external or company internal third parties, information about the transferred data and the form of the data transfer has to be described. Additionally, information about the group of people of the department that has access to the data and is allowed to work with the data has to be given. Consistently, if third parties are involved in data processing and data acquisition, the information about the contractors and clients should be provided. A very important part of the method description is to present a detailed overview of all people that will or could be captured by the cameras. The method description is the second criterion for approval of the camera surveillance request. 
\subsubsection{Video recording device description}
A detailed description for each camera has to be presented in the video recording device document obliged by article 7 and 8 \cite{BayDSG}. The document includes  the cameras' technical details, the camera position at the surveyed area, and the duration of the surveillance. Additionally, it is mandatory to describe the actions taken regarding the camera surveillance labelling requirements based on signboard. In general, the video recording description may overlap with the method description but has its emphasis on the recording itself. The recording description determines the camera position; thus a department group should be careful because a later change of the positions may enforce a new approval request. 
\subsubsection{Data handling description}
Article 7 and 8 \cite{BayDSG} prescribe that the data processing and storage devices have to be described in detail. Also, the software and hardware used for data processes has to be described, as well as action taken for data security regarding data backup, cybercrime, and at due date deletion of the data.
\subsubsection{Staff council letter}
The Bavarian law concerning Personnel Representation(Bayerisches Personalvertretungsgesetz, BayPVG) articles 75a, 76 clause 2(1)(2) \cite{BayPVG} oblige to include the staff council of a company in a camera surveillance approval process. The staff council has to stand up for the rights of the workers of the company they work for. Since the staff cannot evade cameras due to work the staff council has to be asked for approval of the research project in form of an open letter. In contrast, the visitors of a festival are able to avoid the camera because they are informed by the camera surveillance information signboards. 

Except of the staff council letter, all documents have to be handed to the commissioner of data protection of the company. The commissioner will prove the request and release it if it is appropriate.

%% file: Chapter_OnEventSafety.tex
The cameras have to be positioned at high levels. Only high positions of the cameras give an appropriate view on the people and enable proper analysis of the video footage. An obvious problem is how to mount the cameras safely at such a high position. Different solutions are possible, e.g. ladders or similar tools, but, to ensure safety, the use of a lifting platform is mandatory. 

For the usage of lifting platforms, at least two people have to work together: the first controls the platform and the second mounts the cameras. The person controlling the lifting platform needs special training and the platform worker has to be trained in handling the safety gear. Only if both work together and understand each other's commands and signs, the job can be done in a safe way. 

\begin{figure}
\centering
\includegraphics[width=0.5\textwidth]{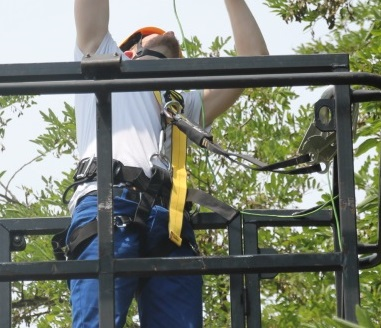}
\caption{\label{fig:Safety1}An example of how the safety gear is used correctly.}
\end{figure}

\begin{figure}
\centering
\includegraphics[width=0.4\textwidth]{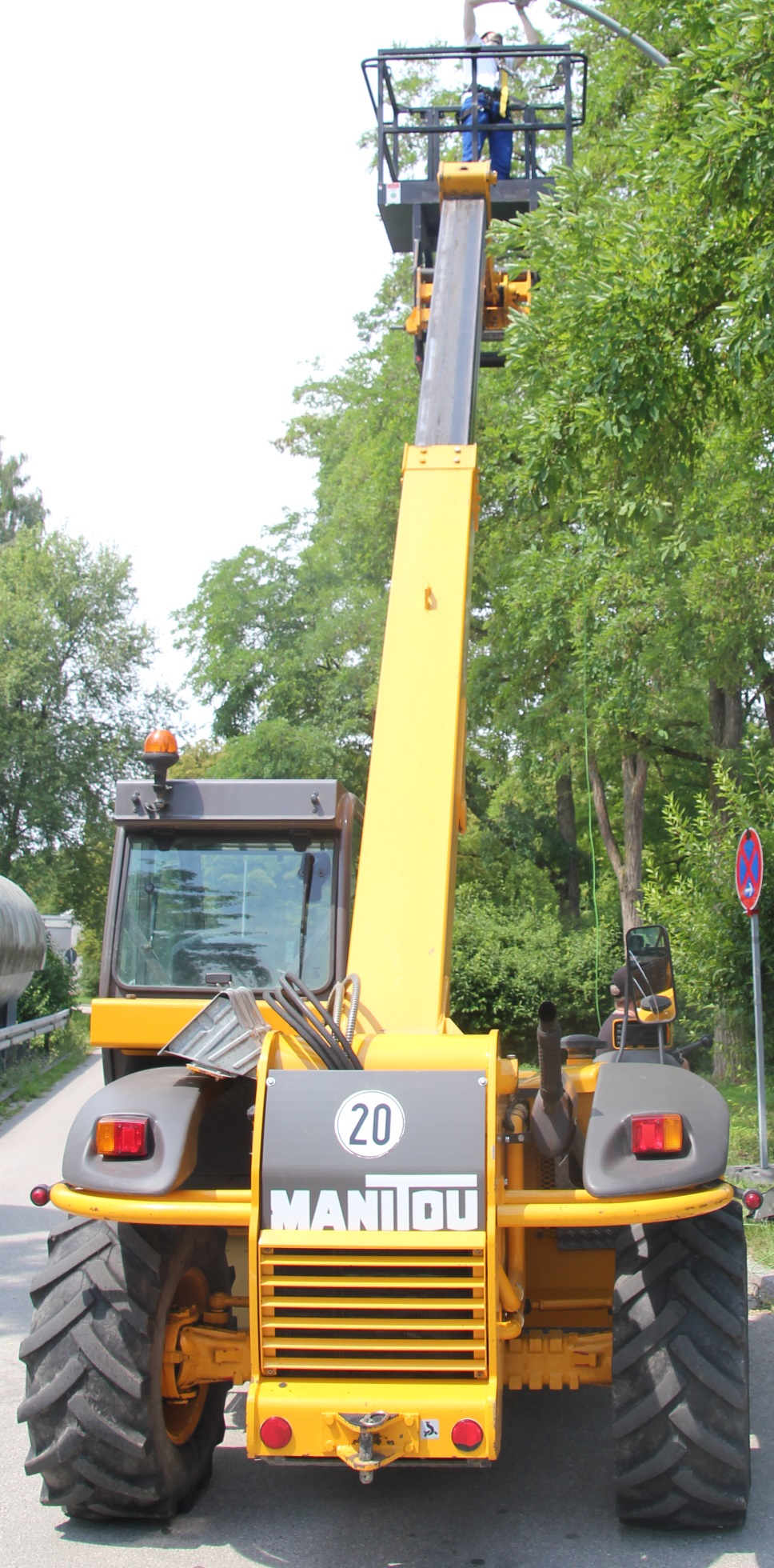}
\caption{\label{fig:Safety2}The lifting platform used by the research team.}
\end{figure}

Employers have to take measures to prevent accidents in accordance with the German Labour Protection Law (Arbeitsschutzgesetz, ArbSchG) \cite{ArbSchG}. Therefore, the employer has to assess the risks of specific work tasks and ensure safety of the workers by instruction, training, and equipment. The German Labour Protection Law applies at its fullest in the case of the camera mounting task by utilising a lifting platform. The person operating on the platform has to be secured by a safety harness, which reduced the risk of injuries due to falling. Safety gear is also mandatory based on the German Association of Occupational Accident Insurance Funds Regulation 1 (DGUV, Vorschrift 1) article 21-31 \cite{DGUV}, which defines regulations for accident prevention. The platform worker also needs additional training regarding correct usage of the safety gear. The training is also required by article 31 \cite{DGUV}. Figure \ref{fig:Safety1} presents a proper example of safety gear usage. In order to comply with article 12 \cite{ArbSchG}, the lifting platform operating worker is also trained for using the device safely. The lifting platform, if incorrectly operated, is not only dangerous for the platform worker, but also for the people at ground level (see Figure \ref{fig:Safety2}). Only well-trained operators should use the lifting platform. 

Aside from the lifting platform based tasks, the research team did not have to consider additional safety precautions.
